Title page

# Prediction of Signal Sequences in Abiotic Stress Inducible Genes from Main Crops by Association Rule Mining

Un-Hyang Ho[1*], Hye-Ok Kong[2]

[1]Faculty of Life Sciences, Kim Il Sung University, Pyongyang

Democratic People's Republic of Korea

[2] Faculty of Mathematics, Kim Il Sung University, Pyongyang

Democratic People's Republic of Korea

*Corresponding author: life1@ryongnamsan.edu.kp;  ryongnam39@yahoo.com

**Abstract**:

**Motivation**: It is important to study on genes affecting to growing environment of main crops. Especially the recognition problem of promoter region, which is the problem to predict whether DNA sequences contain promoter regions or not, is prior to find abiotic stress-inducible genes. Studies on predicting promoter sequences in DNA sequences have been studied by traditional pattern matching methods and machine learning methods in biology and computer science.

**Methods:** We consider the problem of predicting promoter sequences in DNA sequences by using data mining techniques, which extract useful information from large amount of data. We define a new association rule based on hidden values of attributes, construct an association classifier.

**Results**: We perform the experimental evaluation to recognize promoter sequences of abiotic stress inducible genes from rice by using above defined Association Rule based on Hidden Values of Attributes (ARHVA). The classification precision represents the accuracy of promoter recognition using our binary association classifier for a set of DNA sequences of stress inducible genes from rice. And also our classifier provide a much fast and preicise method to predict promoter regions in DNA sequences than former methods in a viewpoint of informatics.

**Availability and implementation:** A web database for the prediction and analysis of promoter is available at http://ppdb.gene.nagoyau.ac.jp/cgi-bin/index.cgi and http://plantpan.mbc.nctu.edu.tw.

**Contact:** life1@ryongnamsan.edu.kp; ryongnam39@yahoo.com

**1 Introduction**

Association Rule Mining (ARM) is a the latest field of data mining study aimed at extracting useful information from large amount of data and using to decision making. Data mining is divided to descripting mining, which provides summary or characteristics about data, and instructive mining, which predicts trends of the future from the past data. ARM is one of the descripting mining techniques producing associations between items in transaction database or relational database.

ARM for classification model is a hybrid method called an association classification (AC), and has been used successfully in many practical applications such as education, medical diagnosis, and web-filtering (Amjad *et al.*, 2009). AC discovers the relations among the values of attributes in records, assigns corresponding classes to them, and makes further precision of the traditional classification methods.

The objects of associations such as itemsets and attribute values are explicitly represented at transactions or records in a given database. In other words, the associations between objects explicitly presented at transactions or records in a database have been considered in former association rule mining.

An ordinal association rule (Campan *et al.*, 2006) which is a variety of association rules, discovered implicit ordinal relations between attributes on the basis of their attribute values. But it was also considered on explicit values of attributes, to discover implicit ordinal relations.

However, associations of data in large databases in many practical applications may be not explicitly represented at records. For example, relations between attributes in a data set consisted of a number of DNA sequences, in which DNA seqences are represented as records and nucleotide bases are represented as attributes, can be extracted from physical and chemical properties of nucleotides. And quantitative values of physical and chemical properties such as molar mass, density and acidity are not explicitly represented in the data set, so that they can be considered as hidden values of attributes.

On the other hand, various meachine learning methods have been studied to recognize biological signal sequences such as promoter which enables transcriptional process (Erratum 2005; Icev 2003; Joseph 1998; Kasabov 2004).

Erratum (2005) provided a hybrid learning system combined explanation based on experimental learning method, and translated knowledge base of relatively pricise domain specific rules to an artificial neural network. And they tested the learning algorithm through promoter identification using a data set of DNA subsequences in 106 *E. coli*, and verified the advantages of their method by comparing with standard back propagation networks, classification trees and nearest neighbor methods. Icev (2003) proposed a new method using neural network to discover signals suggesting existance of promoters in DNA sequences. It was also proposed a new method predicticing promoters in 2111 samples from human and *E. coli* based on the machine learning algorithm called Grey Relational Analysis (GRA) and measured relations between testing sequences and comparative ones in Joseph (1998). Kasabov (2004) developed a new transformation support vector different from the traditional inductive support vector machine and proved its advantage in promoter recognition by training and verifying data sets of 793 vertebrate promoter sequences with 250 bp and 1200 human DNA sequences with the same length.

Studies for gene expression and signal recognition in promoter sequences have been done in not only machine learning but also association rule mining (Kavita 2017, Makihiko 2011).

Kavita (2017) determined expression pattern of certain genes by constructing a new type of association rule based on distance. These rules are based on short DNA sequences called motifs which are contained in promoter and to which certain regulatory proteins are binded. Makihiko (2011) extracted simple association rules and obtained successful results in discovering signals of mammalian promoter sequences.

As above mentioned, the problem of promoter recognition, that is, wheather DNA sequences contain promoter or not, can be discussed on the basis of the relation between nucleotides of which DNA sequences are consisted, and the relations can be extracted by producing additionally chemical and physical characteristics of nucleotides, which are not represented explicitly in a database but existed implicitly.

In this study we define a new association rule based on hidden values of attributes, propose an association classifier, and provide a much fast and pricise method to predict promoter regions in DNA sequences than former methods in a viewpoint of informatics.

## 2 Materials and methods

### 2.1 An association rule based on hidden values of attributes

Let $T = \{t_1, t_2, ..., t_n\}$ be a set of entities (or records) in a database $D$, and $A = \{a_1, a_2, ..., a_m\}$ be a set of attributes in a database $D$, where each entity is characterized by $A$ (a set of attributes).

We denote by $val(t_i, a_j)$ a value of attribute $a_j$ for an entity $t_i$. And we denote by $D_j (j = 1, ..., m)$ a domain of values of attribute $a_j$, which may contain null value. That is, $val(t_i, a_j) \in D_j (i = 1, ..., n)$. Then, we can discuss a relation $r$ (such as $\leq$, $=$ and $\geq$) which is defined (not necessarilly ordinal relation) for $D_{j_1} \times D_{j_2}$. Let $R$ be a set of all possible relations which can be defined for $D_{j_1} \times D_{j_2}$.

Now, we define an Association Rule based on Hidden Values of Attributes (ARHVA) to find various relations between the attributes in a database as followings.

**2.1.1 Definition 1.** We call an association rule $a_{j_1}, a_{j_2}, ..., a_{j_l} \to a_{j_1} r_1 a_{j_2} ... r_{l-1} a_{j_l}$ an *association rule based on hidden values of attributes*, where $\{a_{j_1}, a_{j_2}, ..., a_{j_l}\} \subseteq A$ ($a_{j_u} \neq a_{j_s}$, $u, s = 1, ..., l$, $u \neq s$) and $r_k \in R (k = 1, ..., l-1)$ is the relation for $D_{j_k} \times D_{j_{k+1}}$ ($D_{j_k}$ is a domain of values of an attribute $a_{j_k}$) if

1) $a_{j_1}, a_{j_2}, ..., a_{j_l}$ occur together (not empty) in *sprt* % of $n$ entities. We call this *sprt* the *support* of the rule.

2) $\forall t_{i_0} \in T_r$, $val(t_{i_0}, a_{j_1}) r_1 val(t_{i_0}, a_{j_2}) ... r_{l-1} val(t_{i_0}, a_{j_l})$ where $T_r \subseteq T$ is a set of entities $a_{j_1}, a_{j_2}, ..., a_{j_l}$ occur together. We call $conf = |T_r| / |T|$ the *confidence* of the rule.

Two conditions 1) and 2) are called *support-confidence conditions*.

The length of an association rule based on hidden values of attributes is determined as the number of attributes contained in it. Then, the length of the rule may be possible at most to the total number $m$ of attributes

**2.1.2 Definition 2.** Let denote by *minsprt* and *minconf* the user-defined minimum support and minimum confidence respectively. We call an association rule based on hidden values of attributes *an interesting association rule* if $sprt \geq minsprt$ and $conf \geq minconf$.

### 2.2 An Association classifier for promoter sequences

Given a DNA sequence, an association rule classifier of predcting whether it contains promoter or not, can be constructed by extracting relations between hidden values of arttributes, that is, certain relations between physical and chemical characteristic values of the attributes. Based on this idea, we discuss the problem of recognizing promoter region using the following procedure.

1). Prepare a set of DNA sequences containing promoter, namely, *PT* (a set of positive entities), and a set of DNA sequences not containing promoter, namely, *NT* (a set of negative entities).

2). By using training data, *PT* and *NT*, extract sets of all possible interesting association rules based on hidden values of attributes, *PR* (for positive entity) and *NR* (for negative entity), which satisfy predefined minimum support and minimum confidence.

3). For given a DNA sequence to be classified, if its probability for *PR* is greater than 0.5, it is classified as a positive entity (contains promoter), else as a negative one (not contains promoter).

The above procedure is a working process of an association classifier of predicting promoter sequences, where 1) is a process of preparing training data, 2) is a learning process of the association classifier, and obtained results (*PR* and *NR*) are classification models, and 3) is a process of classification and verification of the association classifier.

In this study, the training data set is a set of of DNA sequences (sequences of A, C, G and T) and their lengths are usually different from each other. However, we focus only on DNA sequences with the fixed length. Because promoter sequences of same functions are combined of the similar motifs, thereby with the fixed length we can consider DNA sequences by setting their transcriptional start points to the same site. Therefore, we prepare *PT* (a set of positive entities) and *NT* (a set of negative entities) consisted of $t_i = a_{i1} a_{i2} .. a_{im}$, $a_{ij} = A\,|\,C\,|\,G\,|\,T$, $(i=1,...,n;\ j=1,...,m)$. That is, we consider sets of DNA sequences with the length-*m*, where *n*, the number of elements in a set, can be different depending on the kinds of training sets.

In order to obtain classification models, *PR* on positive entities and *NR* on negative entities from training sets *PT* and *NT*, we are going to discover only binary association rules with the length-2 rather than any lengths due to structural characteristics of DNA sequences. Binary association rules are in full to predict whether DNA sequences contain promoter or not. If it is an association rule with the greater length than 2, then it is possible to consider binary sub-rules of it, to analyze fully attributes of DNA sequences. Therefore, the training time can be much reduced by searching only binary association rules.

Prior to training it is important to define a set of relations $R = \{r_k,\ k=1,...,l-1\}$ for discovering ARHVAs, where $r_k \in R\,(k=1,...,l-1)$ is a relation between values of attributes reflecting the relation between two nucleotides from DNA sequences(sequences of A, C, G and T). In other words, a binary relation between two nucleotides indicates the relation between certain values reflecting their physical and chemical characteristics. These values are calculated from physical and chemical characteristics (molar mass, melting point, density, ratio of heavy atoms, topological polar surface area, base composition and so on) of nucleotides.

Specified a set of relations, ARHVA mining algorithm is applied to the training sets, namely,

*PT* and *NT*, to extract *PR* (a set of rules for classifying positive entities containing promoter regions) and *NR* (a set of rules for classifying negative entities not containing ones). Therefore, an association classifier operating as a promoter recognizer is composed of *PR* and *NR*.

After training, for a new entity *S* (a DNA sequence to be classified), the following inference is performed by applying it to *PR* and *NR*.

- By using ARHVA mining algorithm, we determine $N_{positive}$ that is the number of ARHVAs satisfying *PR* for *S*. So a number of ARHVAs not satisfying *PR* for *S* is $M_{positive}=|PR|-N_{positive}$.

- By using ARHVA mining algorithm, we determine $N_{negative}$ that is the number of ARHVAs not satisfying *NR* for *S*. So a number of ARHVAs satisfying *NR* for *S* is $M_{negative}=|NR|-N_{negative}$.

- We calculate the probability $P_{positive} = \dfrac{N_{positive} + N_{negative}}{|PR|+|NR|}$ that is the probability for *S* classified as a positive entity.

- We calculate the probability $P_{negative} = \dfrac{M_{negative} + M_{positive}}{|NR|+|PR|}$ or $P_{negative} = 1 - P_{positive}$ that is the probability for *S* classified as a negative entity.

- We classify *S* as a positive entity if $P_{positive} \geq P_{negative}$, else as a negative entity.

**2.3 Promoter prediction of abiotic stress inducible genes from rice (*Oriza sativa Japonica*)**

Here we perform the experimental evaluation to recognize promoter sequences of abiotic stress inducible genes from rice by using above defined ARHVA.

First, we prepare a data set consisited of 100 DNA sequences with 1000 bp length, of which 50 sequences contain promoter regions of stress inducible genes (50 positive entities) and 50 sequences don't contain promoter regions of stress inducible genes (50 negative entities), from rice. From this, for PT and NT, training data sets in this study, $n=50$ and $m=1000$ respectively. This data set can be obtained from Plant Promoter Database (http://ppdb.gene.nagoyau.ac.jp/cgi-bin/index.cgi) and Plant PAN (http://plantpan.mbc.nctu.edu.tw), where each DNA sequence is consisted of 910 bp upstream sequence and 90 bp downstream sequence. Hence a start point of the sequence is -910, and an end point is +90 for the start site transcribed from DNA to RNA.

In our method, binary association rules, that means rules with length-2, are relations between 2 attributes (2 nucleotides) of DNA sequences such as $AC \rightarrow A\,r_{k_1}\,C$, $AG \rightarrow A\,r_{k_2}\,G$, $AT \rightarrow A\,r_{k_3}\,T$ and so on, which satisfy the predefined minimum support and minimum confidence.

To specify a relation set *R* for the set of stress inducible genes from rice, we chose the physical and chemical properties of nucleotides such as molar mass, melting point, density, a ratio of heavy atoms, topological polar surface area and base composition (Table 1).

Table 1. The values reflecting physical and chemical characteristics of nucleotides

| Property | Name of Property | A | C | G | T |
|---|---|---|---|---|---|
| **Property 1** | molar mass | 0.8941 | 0.7351 | 1.0000 | 0.8344 |
| **Property 2** | melting point | 1.0000 | 0.8897 | 0.9931 | 0.9269 |
| **Property 3** | Density | 0.7272 | 0.7045 | 1.0000 | 0.5590 |
| **Property 4** | a ratio of heavy atoms | 1.0000 | 0.6734 | 0.8609 | 0.8156 |
| **Property 5** | topological polar surface area | 0.8367 | 0.7016 | 1.0000 | 0.6049 |
| **Property 6** | base composition | 1.0000 | 0.7592 | 0.7606 | 0.9999 |

In Table 1, for each property, the values of nucleotides are calculated by normalizing the values corresponding physical and chemical properties. For example, property 4, a ratio of heavy atoms, means a ratio of the numbers of C, N and O atoms in each nucleotide, is calculated by normalizing A(6:4:0), C(4:3:1), G(5:5:1) and T(5:2:2), while property 6, base composition, is the value calculated and normalized base composition for each nucleotide in whole genome of *Oryza sativa Japonica* in GenBank database.

On account of the purpose of finding the relations between 4 nucleotides based on the values of physical and chemical properties by applying ARHVA, we found property 2 and property 4 are equal, similarly property 3 and property 5 are equal in Table 1. Therefore, the only relations between property 1 (molar mass) and property 6 (base composition), could be selected for binary classification whether DNA sequence contains promoter region or not. The relation could be =, $<_1$, $<_6$, $>_1$, and $>_6$, where subscripts represent numbers of the properties in Table 1. In the binary classification, in case of DNA sequence containing promoter a value of target classification is 1, and in the case of DNA sequence not containing promoter a value of target classification is 0.

Using an absolute value of Spearman's rank correlation coefficient (Pawel, 2015), we evaluated the correlations between nucleotide attributes and target classification attributes, with the result that property 1 and property 6 correlated considerably to the targets (the average correlations of property 1 and property 6 are 0.0083 and 0.0341 respectively, which are over the mean value).

After defined the set of relations *R*, we preprocessed the training data sets, *PT* and *NT*. Because the set of relations *R* is defined from property 1 and property 6, in preprocessing we ignored the attributes with the correlations less than a certain threshold, that is defined as a value near by the minimum correlation, by considering the correlations between the attributes and the target values, and made the precision of the classification higher. Figure 1 and Figure 2 show Spearman's rank correlation coefficients between property 1 / property 6 and target attribute respectively, for the sequences with 100 bp among promoter sequences (1000 bp) of stress inducible genes from rice. Figure 3 shows overlapping of the correlations for property 1 and property 6, and it is a pattern representing the relations between the attributes of DNA sequences, at which we aimed.

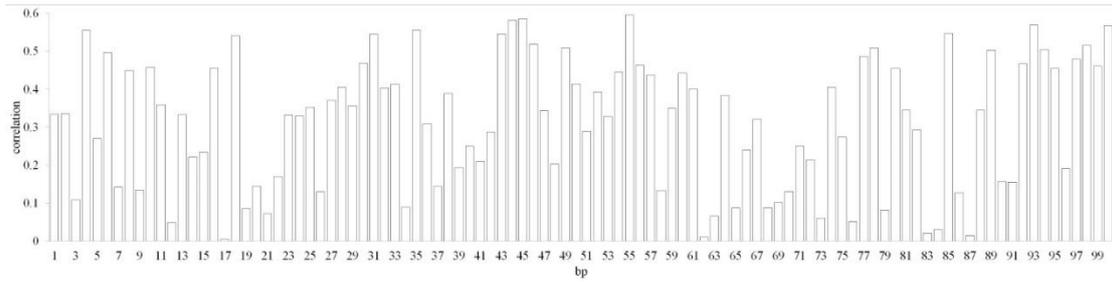

Figure 1. Spearman's rank correlation of property 1 for promoter sequence (100 bp) of stress inducible genes from rice

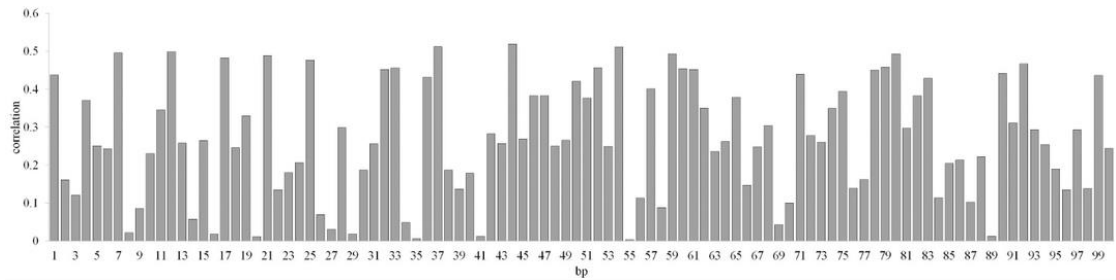

Figure 2. Spearman's rank correlation of property 6 for promoter sequence (100 bp) of stress inducible genes from rice

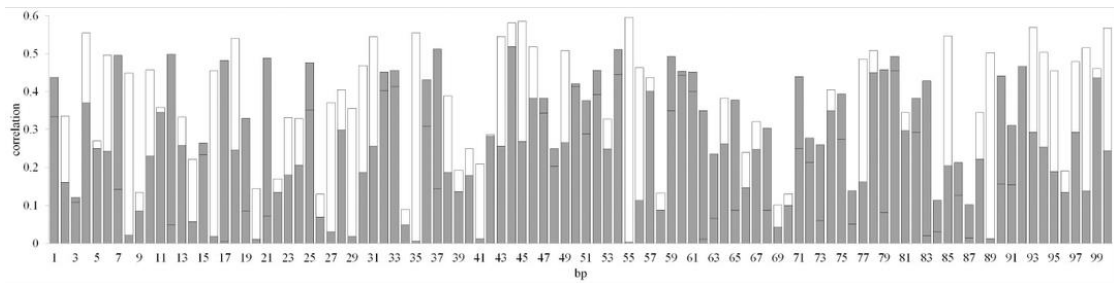

Figure 3. Overlapping of Spearman's rank correlations of property 1 and property 6

Thus a set of the relations for *PT* and *NT*, sets of DNA sequences of stress inducible genes from rice, was defined as a set { =, $<_1$, $<_6$, $>_1$, $>_6$} of orderal relations for characteristic values of molar mass (property 1) and base composition (property 6). From this, we extracted *PR* and *NR*, setting the minimum support at 0.95, and decreasing the minimum confidence by 0.05 from minconf=0.75 to minconf=o.4, and produced an association classifier of recognizing promoter regions.

By using the produced an association classifier, results by promoter classification for DNA sequences are shown in Table 2.

In Table 2, the classification precision represents the accuracy of promoter recognition using our binary association classifier for a set of DNA sequences of stress inducible genes from rice. As shown in Table 2, it shows trend that the more confidence threshold is decreased, the more the precision is increased. And the more the confidence threshold is decreased, the more classification speed is high, and the speed is a very fast speed comparing with the traditional searching methods.

Table 2. Experimental results according to the minimum confidence

| minimum | classification | number of false | number of false |
|---|---|---|---|
| **0.75** | 0.91 | 3 | 6 |
| **0.70** | 0.91 | 5 | 4 |
| **0.65** | 0.92 | 3 | 5 |
| **0.60** | 0.92 | 4 | 4 |
| **0.55** | 0.95 | 3 | 2 |
| **0.50** | 0.94 | 3 | 3 |
| **0.45** | 0.97 | 2 | 1 |
| **0.40** | 0.98 | 1 | 1 |

**3. Results and Discussion**

In this paper, we proposed a classification model using association rule mining based on hidden values of attributes for promoter prediction.

So far association rule mining (ARM) (Kavita, 2017) has been discussed the relations between objects by using the items or the values of attributes existing explicitly in a database, while they has also discussed ordinal relations that existed implicitly, by using the values of attributes that existed explicitly, in ordinal association rule (Campan *et al.*, 2006). In our association rule mining based on hidden values of attributes (ARHVA), the relations between objects can be revealed by using the values of attributes existing implicitly in a database.

Most biological signal sequence recognizer using statistical methods (Joseph and Winston, 1998; Erratum, 2005; Makihiko, 2011) and machine learning methods (Towell *et al.*, 1990; Pedersen and Engelbrecht 1995; Tatavarthi 2011; Kasabov and Pang, 2004; Makihiko 2011, 2012; Wang *et al.*, 2007; Amjad *et al.*, 2009) may identify signal sequences in DNA sequences by using searching methods based on pattern matching. However, in our promoter classifier, the existence of signal sequences is determined first of all, so needless searching is disappeared.

Biological signal recognizers using association rules (Icev *et al.*, 2003; Shibayama *et al.*, 1995) can recognize only a short DNA sequences such as motifs in promoter regions, but our classifier processes whole sequences of promoters.

In this paper, we constructed a promoter classifier of reflecting the structural characteristics of DNA sequences on the basis of ARHVA mining. Especially, by considering only binary association rules, we made analysis for DNA sequences very effect, and confirmed the validity through experiments.

The method proposed in this paper is a pattern recognition one using recent data mining techniques in bio-genome databases of huge amount, being more efficient rather than the traditional searching methods.

# References

1. Amjad, A. et al. (2009) N4: A precise and highly sensitive promoter predictor using neural network fed by nearest neighbors. *Genes & Genetic Systems*, 84, 425-430.
2. Campan, A. et al. (2006) An algorithm for the discovery of arbitrary length ordinal association rules. The 2006 International Conference on Data Mining (DMIN06), Las Vegas, USA, pp 107-113.
3. Erratum, (2005) Human pol II promoter prediction: time series descriptors and machine learning. *Nucleic Acids Research*, 33(4), 1332-1336.
4. Icev, A. et al. (2003) Distance-based association rules mining. M. J. Zaki, J. T.-L. Wang, and H. Toivonen (Eds.), Proceedings of the 3rd ACM SIGKDD Workshop on Data Mining in Bioinformatics (BIOKDD'03), August 27, 2003, Washington DC, pp. 34–40.
5. Joseph, O. and Winston, H. (1998) A Statistical Model for Prokaryotic Promoter Prediction. *Genome Informatics*, 9, 271-273.
6. Kasabov, N. and Pang, S. (2004) Transductive support vector machines and applications in bioinformatics for promoter recognition. *Neural Information Processing Letters and Reviews*, 3(2), 31-37.
7. Kavita, M. and Gaurav, A. (2017) A comparative study of association rule mining techniques and predictive mining approaches for association classification. *International Journal of Advanced Research in Computer Science*, 8(9), 365-372.
8. Makihiko, S. (2011) GC Wave Analysis in Promoter Regions via Wavelet Analysis and Support Vector Machine. *Procedia Computer Science*, 6, 285-290.
9. Makihiko, S. (2012) Promoter Analysis with Wavelets and Support Vector Machines. *Procedia Computer Science*, 12, 432-437.
10. Pawel, C. (2015) *Data Mining Algorithms: Explained Using R.* WILEY, pp 321-324.
11. Pedersen, A. and Engelbrecht, J. (1995) Investigations of *Escherichia coli* promoter sequences with artificial neural networks: New signals discovered upstream of the transcriptional startpoint. *Proc Int Conf Intell Syst Mol Biol.* 3, 292-299.
12. Shibayama, G. et al. (1995) Mining association rules from signals found in mammalian promoter sequences. 6, 108–109.
13. Tatavarthi, UD. et al. (2011) Padmanbhuni VNR, Allam AR, Gumpeny RS.; In silico promoter prediction using grey relational analysis. *Journal of Theoretical and Applied Information Technology*, 24(2), 107-112.
14. Towell, GG. et al. (1990) Refinement of approximate domain theories by knowledge-based artificial neural networks. *In: Proceedings of the Eighth National Conference on Artificial Intelligence* (AAAI-90), 861-866.
15. Wang, J. et al. (2007) MetaProm: a neural network based meta-predictor for alternative human promoter prediction. *BMC Genomics*, 17(8), 374.

**Figure legends**

Figure 1. Spearman's rank correlation of property 1 for promoter sequence (100 bp) of stress inducible genes from rice

Figure 2. Spearman's rank correlation of property 6 for promoter sequence (100 bp) of stress inducible genes from rice

Figure 3. Overlapping of Spearman's rank correlations of property 1 and property 6